\documentclass[12pt]{article}
\usepackage{amssymb}
\usepackage[T1]{fontenc}
\usepackage{subcaption,graphicx}
\usepackage[numbers]{natbib}
\usepackage{mathtools}  
\usepackage{amssymb}    
\usepackage{amsthm}
\usepackage{subfiles}
\usepackage{hyperref}
\hypersetup{
  hidelinks, 
}
\usepackage{booktabs}
\usepackage{caption}
\usepackage{subcaption}
\usepackage{tabularx}
\usepackage{graphicx}
\usepackage{multirow}
\usepackage{array}
\usepackage{microtype}
\usepackage{array,tabularx}

\begin{document}

\begin{titlepage}
\raggedright

{\Large\bfseries Comparative Analysis of Global and Local Probabilistic Time Series Forecasting for Contiguous Spatial Demand Regions\par}

\vspace{1.25cm}

{\bfseries Jiahe Ling, M.S. (Corresponding Author)}\par
Department of Statistics\par
University of Chicago\par
5747 South Ellis Avenue, Chicago, IL 60637\par
Tel: 608-977-4956; Email: \texttt{sling@uchicago.edu}\par

\vspace{0.9cm}

{\bfseries Wei-Biao Wu, Ph.D.}\par
Department of Statistics\par
University of Chicago\par
5747 South Ellis Avenue, Chicago, IL 60637\par
Tel: 773-702-0958; Email: \texttt{wbwu@uchicago.edu}\par

\vspace{0.9cm}

\vfill

{\bfseries Author Contributions}\par
The authors confirm their contributions to the paper as follows: study conception and design: J. Ling, W. B. Wu; data collection: J. Ling; analysis and interpretation of results: J. Ling; draft manuscript preparation: J. Ling, W. B. Wu. All authors reviewed the results and approved the final version of the manuscript.

\end{titlepage}

\begin{abstract}

This study evaluates three probabilistic forecasting strategies using LightGBM: global pooling, cluster-level pooling, and station-level modeling across a range of scenarios, from fully homogeneous simulated data to highly heterogeneous real-world Divvy bike-share demand observed during 2023 to 2024. Clustering was performed using the K-means algorithm applied to principal component analysis transformed covariates, which included time series features, counts of nearby transportation infrastructure, and local demographic characteristics. Forecasting performance was assessed using prediction interval coverage probability (PICP), normalized interval width (PINAW), and the mean squared error (MSE) of the median forecast. The results show that global LightGBM models incorporating station identifiers consistently outperform both cluster-level and station-level models across most scenarios. These global models effectively leverage the full cross-sectional dataset while enabling local adjustments through the station identifier, resulting in superior prediction interval coverage, sharper intervals, and lower forecast errors. In contrast, cluster-based models often suffer from residual within group heterogeneity, leading to degraded accuracy. Station-level models capture fine-grained local dynamics in heterogeneous settings. These findings underscore that global LightGBM models with embedded station identifiers provide a robust, scalable, and computationally efficient framework for transportation demand forecasting. By balancing global structure with local specificity, this approach offers a practical and effective solution for real-world mobility applications.

\end{abstract}

\newpage
\vspace{1.5in}
\tableofcontents

\newpage

\section{Introduction}

Demand forecasting in spatially contiguous regions underpins critical decisions in transportation, logistics, and urban planning, where accurate prediction of future usage distributions can substantially improve resource allocation and service quality. For example, DoorDash uses spatio-temporal demand forecasts to match Dashers with incoming orders in each neighborhood, ensuring drivers are in the right place at the right time and reducing wait times during sudden surges from events or weather changes \citep{sajin_managing_2021}. Uber similarly employs machine learning and probabilistic forecasting models to anticipate rider demand at the neighborhood level, dynamically dispatching drivers and adjusting pricing to reduce response times and avoid both oversupply and undersupply of vehicles \citep{zhu_deep_2017}. These industry practices underscore the need to balance rich models with practical computational constraints, setting the stage for our exploration of contiguous spatial demand clusters as a means to harness pooled forecasts efficiently in real-world settings. Despite their power, these data-rich systems must still reckon with limits on training time and compute resources, especially when scaling to thousands of micro targets.

Unpooled local models fit separate forecasting functions to each individual time series, capturing idiosyncratic patterns at the cost of a model count that grows linearly with the number of series. In contrast, fully pooled global models merge data from all series into a single training set, gaining statistical efficiency and maintaining constant model complexity regardless of series count. Hybrid clustering approaches strike a middle ground by grouping series into homogeneous clusters before applying global modeling within each cluster, thus blending information sharing with localized adaptability. Importantly, global forecasting methods can match local methods without any series similarity assumptions and that, by applying generalization bounds based on multitask learning, local complexity grows with the number of series while global complexity stays constant so global models generalize better on large datasets \citep{montero-manso_principles_2021}. The main challenge of applying such fine-grained global models is their high computational cost, which industry leaders like DoorDash address by grouping delivery zones into contiguous micro regions for pooled forecasting \citep{sajin_managing_2021}.

Probabilistic forecasting quantifies uncertainty by providing a full distribution of possible outcomes rather than a single point estimate. By explicitly modeling demand variability, organizations can allocate resources more efficiently and reduce waste in inventory and staffing. Moreover, in risk-sensitive contexts such as supply chains and on-demand services, probabilistic forecasts enable decision makers to anticipate extreme scenarios and develop contingency plans \citep{sun_optimal_2023}. However, despite extensive advances in probabilistic forecasting and the emergence of unpooled local models, fully pooled global models, and hybrid clustered approaches on time series forecasting, no study has yet systematically compared how these differing pooling paradigms perform in producing probabilistic forecasts.

To address this gap, the present study conducts a rigorous comparison of pooling strategies using gradient boosting as implemented in LightGBM. Homogeneous synthetic datasets are generated by simulating each station’s series from four data generating processes whose parameters are estimated on the pooled average demand series, while the real Divvy bicycle share data provide a heterogeneous benchmark. Under three information sharing schemes (fully pooled global estimation, unpooled station-specific estimation, and clustered estimation), LightGBM is trained on each synthetic and real dataset, and its probabilistic forecasts are evaluated. The results reveal the conditions under which global pooling, local modeling, or clustered pooling deliver the best trade-off between forecast distributional richness and training efficiency, and thus inform the deployment of practical probabilistic forecasting systems in transportation and logistics.

The remainder of this paper is organized as follows. The literature review section presents a comprehensive review of relevant studies, starting with probabilistic forecasting methods for time series, then exploring unpooled local, fully pooled global, and hybrid clustered approaches, and concluding with an overview of time series clustering techniques. The data section describes the data used in this study, covering the Chicago Divvy bicycle-sharing records, public transportation information, the geographic and socioeconomic covariates, and the simulated demand series generated under both homogeneous and heterogeneous regimes. The methodology section details the methodology for model fitting and evaluation, outlining how each forecasting model is implemented within the three pooling paradigms and how probabilistic forecasts are produced and assessed. The results and discussion section presents the empirical results from both synthetic and real-world experiments. Finally, the conclusion section summarizes the key findings, discusses practical implications for demand forecasting in transportation and logistics, acknowledges the limitations of this research, and suggests directions for future work. By systematically comparing pooling strategies under varied demand dynamics, this study offers essential guidance for designing probabilistic forecasting systems that balance expressiveness with computational feasibility, ultimately enhancing decision-making in on-demand services and urban planning.

\newpage

\section{Literature Review}

\subsection{Time Series Probabilistic Forecasting}

The classical forecasting methods, including Autoregressive integrated moving average (ARIMA), exponential‐smoothing and Theta methods, derive prediction intervals by first casting the forecasting model into an equivalent state‐space or infinite moving average (MA) form and then either analytically computing or empirically approximating the forecast error variance. ARIMA models construct prediction intervals by rewriting the fitted model in its infinite MA form to quantify the variance of the forecast error from the innovation variance and impulse response coefficients, and under a Gaussian error assumption, the appropriate normal quantile is then applied to this variance to generate symmetric confidence intervals around the point forecasts \citep{chatfield_calculating_1993}. Similarly, exponential‐smoothing methods, like Simple Exponential Smoothing (SES) and Holt’s linear and Holt–Winters seasonal methods, admit an equivalent Gaussian state‐space form in which the impulse‐response coefficients determine the forecast error variance under additive errors, and by applying the appropriate normal quantile to the square root of this variance one obtains analytic symmetric confidence intervals, whereas multiplicative or nonlinear variants employ Monte Carlo simulation of the state‐space recursion and use empirical quantiles of the simulated forecast ensemble to define prediction bounds \citep{hyndman_state_2002}. The Theta method also admits an additive representation of error level and drift state space, from which forecast error variances are obtained analytically via its infinite MA form and symmetric Gaussian-based prediction intervals follow by applying the appropriate normal quantile \citep{assimakopoulos_theta_2000}. For nonlinear extensions, predictive distributions are generated by Monte Carlo simulation of the state-space recursion, with interval bounds of a $(1-\alpha)100\%$ interval defined by the empirical $\alpha/2$ and $1-\alpha/2$ quantiles of the simulated ensemble \citep{assimakopoulos_theta_2000}. The Quantile Regression (QR) constructs prediction intervals by directly estimating the conditional quantile functions through minimization of an asymmetric absolute-deviation loss at the desired levels, so that fitting regressions at $\alpha/2$ and $1-\alpha/2$ yields the lower and upper bounds of a $(1-\alpha)100\%$ interval, which accommodates heteroscedasticity and skewed distributions by making no Gaussian error assumptions \citep{koenker_regression_1978}.\par

On the other hand, machine learning methods directly model the conditional distribution or its quantiles using techniques such as quantile regression, ensemble methods, Bayesian inference, and diffusion models to generate prediction intervals. Quantile Regression Forests (QRF) extend this idea by using the terminal node memberships of each tree to assign adaptive weights to training observations and averaging these weights across the ensemble to form an empirical conditional Cumulative Distribution Function (CDF) at the new input, and then the bounds of a $(1-\alpha)100\%$ interval are obtained by reading the quantiles $\alpha/2$ and $1-\alpha/2$ of this weighted distribution \citep{meinshausen_quantile_2006}.  Gradient Boosting Machines (GBM) can also yield full predictive distributions. For example, NGBoost treats the parameters of a chosen output distribution such as the mean and variance of a Gaussian as functions of the inputs and sequentially fitting them by multiparameter boosting using natural gradients to correct parameterization, thus producing a full predictive distribution from which confidence intervals are obtained either by applying the appropriate normal quantiles or by extracting empirical quantiles of the fitted distribution \citep{duan_ngboost_2020}. To obtain prediction intervals from Support Vector Regression (SVR), Bayesian SVR is commonly used to construct prediction intervals by casting the $\epsilon$ insensitive loss within a Bayesian framework, employing a Laplace approximation around the MAP solution to derive a predictive variance from kernel covariance, augmenting this with an observation noise term, and then applying Gaussian quantile bounds around the point forecast \citep{gao_probabilistic_2002}. Finally, Gaussian processes (GP) for regression derive exact posterior predictive distributions by conditioning the joint Gaussian prior to training data to obtain a Gaussian forecast with analytically computed mean and variance, and then construct confidence intervals by applying the appropriate normal quantile to the predictive standard deviation \citep{williams_gaussian_1995}. \par

Complementing these machine learning techniques, several specialized methods have been developed to equip neural networks with calibrated predictive uncertainty. The Delta method constructs confidence intervals by linearizing the neural network around its fitted parameters, propagating parameter uncertainty through this local approximation to estimate the variance of the forecast, and then using a normality assumption to define symmetric intervals around the point prediction \citep{kollovieh_predict_2023}. Building on this idea of parameter uncertainty propagation, mean–variance estimation (MVE) augments a standard neural network with a second output unit that predicts the conditional variance of the target under an assumed error distribution and trains both outputs jointly through a likelihood-based loss, with the resulting variance head providing input-dependent uncertainty estimates for the construction of confidence intervals \citep{nix_estimating_1994}. Alternatively, bootstrap methods construct confidence intervals by training an ensemble of neural networks on multiple bootstrap-resampled versions of the training data and then using the empirical variability of their predictions to quantify model uncertainty, with interval bounds derived from the spread of the ensemble outputs \citep{carney_confidence_1999}. From a Bayesian perspective, neural networks quantify predictive uncertainty by placing a prior over their weights and performing approximate posterior inference via variational methods or Monte Carlo dropout, drawing multiple stochastic weight samples at prediction time to estimate epistemic uncertainty, combining this with an aleatoric variance term to obtain the total predictive variance, and constructing confidence intervals around the posterior mean forecast by applying the appropriate normal quantiles to this variance \citep{zhu_deep_2017}. \par

Beyond these neural network-specific approaches, diffusion models have recently been applied to probabilistic prediction by using an unconditional denoising diffusion model trained on complete time series and then steering the reverse-diffusion process at inference to enforce consistency with observed history and generate an ensemble of future trajectories, from which empirical confidence intervals are derived \citep{kollovieh_predict_2023}. For example, TSDiff employs Bayes-derived adjustment terms based on mean-square or quantile losses against observed data to guide each reverse-diffusion sample toward the target distribution, enabling calibrated prediction intervals either via empirical percentile computation or through direct quantile self-guidance under an asymmetric Laplace formulation \citep{kollovieh_predict_2023}.

\subsection{Local, Global, and Hybrid Time Series Forecasting Methods}

Time series forecasting is critical across many domains, particularly in demand forecasting and inventory planning. Several studies have highlighted its significance in retail and supply chain applications \citep{sun_planning_2022, cheng_research_2024, yin_understanding_2023}. A fundamental modeling choice is whether to use unpooled local models or pooled global models. Local models are trained independently on each series, treating each as an isolated prediction task. In contrast, global models fit a single forecasting function in a series collection, sharing information among them \citep{yingjie_local_2024}. By fitting each series independently, local models can capture idiosyncratic patterns with low bias but often suffer high variance when data are limited; on the other hand, global models enforce a shared set of patterns across all series, which may introduce bias by smoothing over series-specific nuances but substantially reduce estimation variance by leveraging the entire dataset \citep{montero-manso_principles_2021}. \par

Montero-Manso and Hyndman (2021) establish that any forecast obtained by fitting separate local models to each series can be equivalently obtained by a single pooled global model, illustrating that there is no loss of expressiveness in adopting a global approach \citep{montero-manso_principles_2021}. They further derive generalization bounds showing that the complexity penalty for a collection of local models grows with the product of their individual hypothesis class sizes, whereas the penalty for a global model depends only on its overall hypothesis class size. As a result, sufficiently rich global models enjoy strictly tighter worst-case bounds than their local counterparts. Building on these theoretical insights, Hewamalage et al. (2022) conduct Monte Carlo experiments on both homogeneous and heterogeneous collections of time series, simulating simple linear processes, namely an Autoregressive model of order 3 (AR(3)) and a Seasonal Autoregressive model of order 1 (SAR(1)), as well as progressively more complex nonlinear processes, namely the Chaotic Logistic Map, the Self‐Exciting Threshold Autoregressive (SETAR) model, and the Mackey‐Glass delay differential equation, across 100 or 1000 replicates \citep{hewamalage_global_2022}. They compare local methods with linear and nonlinear global approaches, which include Feed‐Forward Neural Networks (FFNN), Recurrent Neural Networks (RNN), and Light Gradient Boosting Machines (LightGBM). Forecast accuracy is assessed using Mean Absolute Scaled Error (MASE) and Symmetric Mean Absolute Percentage Error (SMAPE). Their results indicate that when series are long and homogeneous, correctly specified local models, particularly simple linear ones, outperform all global alternatives. Conversely, when series are short or data are heterogeneous, nonlinear global models, especially RNNs and LGBMs, achieve lower errors than both linear global and local methods. Linear global models perform well in multiple short series settings but require extensive feature engineering to remain competitive under heterogeneity, and incorporating a known group indicator enhances performance without fully matching the accuracy of ideal local models.

Wellens et al. (2023) generate synthetic datasets from AR(1) and AR(5) processes with controlled heterogeneity in intercepts and autoregressive coefficients and vary both the number of observations per series and the degree of cross-series variability, and in the study a weekly retail sales dataset from the real-world featuring series of different lengths and seasonal patterns is also included \citep{wellens_when_2023}. Global methods comprise AR with cross‐series dummy variables for seasonality and LightGBM models. These are compared with local methods in which AR or LightGBM are fit separately to each series. Forecast accuracy is assessed using the root mean squared scaled error (RMSSE), while bias is measured using the scaled mean error (SME) and its absolute counterpart (MASE). Wellens et al. (2023) conclude that heterogeneity makes it difficult for global methods to estimate parameters well, so that well‐specified local methods usually win when there are ample data and lags available, but when data are scarce, global methods tend to outperform \citep{wellens_when_2023}. Moreover, non-linear global methods naturally handle heterogeneity better than linear ones, making non-linear globals the preferable choice in most real-world scenarios. \par

Hybrid strategies that combine global and local components have also been developed. Sen et al. (2019) propose DeepGLO, a hybrid framework that integrates a regularized low-rank matrix factorization with a Temporal Convolutional Network to capture shared temporal patterns globally and then refines each series' forecast locally through a second Temporal Convolutional Network incorporating the series’s recent history, covariates, and the global forecast \citep{sen_think_2019}. DeepGLO eliminates the need for manual normalization and consistently surpasses both purely global and purely local baselines on high-dimensional forecasting benchmarks. In addition to leveraging global information to inform the construction of local models, Bandara et al. (2020) group time series into homogeneous groups based on interpretable characteristics such as seasonality strength and autocorrelation using methods such as K-Means or DBSCAN \citep{bandara_forecasting_2020}. A separate LSTM model is trained for each cluster where inputs are deseasonalized via Seasonal Decomposition of Time (STL) and normalized over sliding windows before being fed into a multi-step ahead configuration. These clustered LSTM models consistently outperform a single global LSTM and achieve lower SMAPE and MASE compared to the leading statistical and machine learning baselines.

In summary, previous work has shown that the choice between local and global forecasting models depends on data availability, homogeneity, and model complexity, with nonlinear global approaches often excelling under heterogeneity, while well-specified local methods prevail when there is a large amount of homogeneous data. However, these studies have largely focused on point forecasts and deterministic error metrics, leaving a gap in understanding how probabilistic forecasting behaves under pooled versus un-pooled paradigms. Thus, this study aims to address this gap by comparing the performance of global and local probabilistic models within spatially contiguous demand clusters.

\subsection{Time Series Clustering} 

Time series clustering methods are typically classified into three principal categories: shape based, feature based, and model based clustering \citep{aghabozorgi_time-series_2015}. Firstly, shape based clustering methods operate directly on raw time series by employing elastic similarity measures such as dynamic time warping, longest common subsequence, or normalized cross correlation to align sequences and quantify dissimilarity. Once a pairwise distance matrix is obtained, conventional grouping algorithms such as agglomerative hierarchical clustering, k medoids, or density based clustering are applied to cluster series whose waveforms exhibit similar temporal structures \citep{warren_liao_clustering_2005}. However, the pairwise distance computations incur quadratic time complexity and can be sensitive to noise and outliers unless supplemented by smoothing or lower bounding techniques \citep{warren_liao_clustering_2005}.

In contrast, feature-based techniques begin by transforming each series into a fixed-length vector of summary statistics and then apply vector space clustering methods such as K-means \citep{bandara_forecasting_2020}. Commonly used features include central moments (mean, variance, skewness, kurtosis), autocorrelation coefficients at multiple lags, measures of trend and seasonal strength, spectral features (dominant frequencies, spectral entropy), summary stationarity metrics, counts or distances of discriminative subsequences, and transform-based coefficients (wavelet or Fourier coefficients) \citep{warren_liao_clustering_2005}. These methods reduce each series to a fixed-length feature vector, which facilitates interpretability and accommodates series of varying lengths, but their success depends critically on the relevance and quality of the extracted features.

Finally, model based clustering assumes each series $y_i(t)$ arises from one of $K$ parametric models with prior weights $\pi_k$, fits this finite mixture via the expectation maximization algorithm to estimate both model parameters $\{\theta_k\}$ and posterior probabilities $\gamma_{ik}$, and then assigns each series to the cluster with the highest $\gamma_{ik}$, thereby grouping series by shared temporal dynamics \citep{chamroukhi_model-based_2011}. The primary limitations of the model based approach are the computational cost of per-series model fitting and sensitivity to model misspecification, particularly for short or irregular series.

More recently, deep learning for time series clustering uses neural network models such as convolutional autoencoders or recurrent autoencoders to encode each sequence into a low dimensional vector and decode it to reconstruct the input, ensuring representation quality \citep{alqahtani_deep_2021}. A clustering loss, for example K-means or a mixture model objective, is combined with the reconstruction loss and optimized in an end-to-end training so that representation and grouping inform each other. After training, each series is assigned to the group whose prototype or component best matches its vector. These methods handle noise, complex dynamics, and high dimensional data more robustly than approaches that separate embedding from grouping. However, these methods demand substantial computational resources and large volumes of training data, and the performance is sensitive to network architecture and hyperparameter choices, and the resulting latent representations often lack interpretability.

\newpage

\section{Data}

\subsection{Chicago Divvy Bicycle Sharing Data}
\label{sec:divvydata}

Ridership data are sourced from the City of Chicago’s Divvy Open Data portal, covering every recorded trip between April 1, 2020 and May 31, 2025 \citep{divvy_divvy_2025}. This study restricts the analysis to recorded trips between January 1, 2023 and December 31, 2024, because data from 2020 to 2022 are influenced by the COVID-19 pandemic and 2025 records remain incomplete. This study analyzes 11,580,445 Divvy trip records collected between January 2023 and December 2024, originally spanning 1,809 unique stations. To align with Chicago’s official 50-ward framework, the stations were spatially joined to the 2023 ward boundaries: 1,782 stations fell within these limits, while 27 peripheral stations were excluded, reducing the trip count to 11,516,339 at raw resolution. 

Because trips in bike sharing systems exhibit strong intraday patterns, including morning and evening peaks, lunchtime spikes, and off-peak lulls that are obscured at coarser temporal scales, this study aggregates counts into hourly intervals to balance temporal detail and statistical stability, avoiding the sparsity of minute level data and the loss of critical rush hour dynamics in daily or weekly aggregates. Torres et al. (2024) demonstrate that the addition of trip counts in hourly intervals or over multiple hours significantly reduces statistical noise while preserving critical demand patterns, and this approach reflects common practice in the literature, for example, in the study by Hosseini and Hosseini (2020) \citep{torres_forecasting_2024, hosseini_model_2020}. Consequently, this study adopts hourly bins as its temporal aggregation level to capture essential intraday demand patterns while minimizing statistical noise. Aggregating into an hourly time series by station produces 3,553,604 stations' hourly observations of 1,782 stations. To ensure a complete 24-hour profile for each station, any hour with no recorded trips was imputed with a demand of zero. 

\begin{figure}[htbp]
  \centering
  \includegraphics[width=0.8\textwidth]{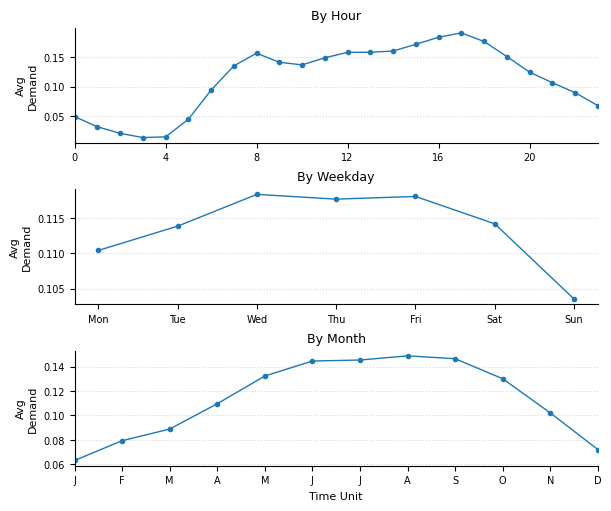}
  \caption{Temporal demand profiles by hour of day, day of week, and month of year}
  \label{fig:trend}
\end{figure}

In \autoref{fig:trend}, the daily profile exhibits a morning and late-afternoon peak, the weekly profile shows midweek elevation, and the monthly profile reveals higher demand in summer with a decline in winter. \autoref{tab:hourly_stacked_summary} shows that hourly counts are evenly distributed on Monday through Saturday, with a modest reduction on Sunday, indicating consistent usage on weekdays and weekends. Seasonality peaks in summer, followed by spring and fall, with winter accounting for the fewest observations.

To investigate whether time trends are heterogeneous across stations, suppose each station’s hourly demand $y_{it}$ is modeled as $y_{it}=\alpha_i+\beta_i t+\varepsilon_{it}$, where $\alpha_i$ is the station‐specific intercept, $\beta_i$ the per‐hour trend slope, and $\varepsilon_{it}\sim(0,\sigma_i^2)$, and the null hypothesis $H_0:\beta_1=\cdots=\beta_N$ of a common slope is tested against heterogeneous alternatives using the bias‐adjusted $\widetilde D$ statistic of \cite{hashem_pesaran_testing_2008}. First, the within‐transformation $M=I_T-T^{-1}\mathbf{1}_T\mathbf{1}_T^\top$, where $T$ denotes the length of each station’s hourly demand series, is applied to purge station intercepts. Station‐level estimates are obtained by $\hat\theta_i=(X_i^\top M X_i)^{-1}X_i^\top M y_i$ and $\hat\sigma_i^2=(T-2)^{-1}(y_i-X_i\hat\theta_i)^\top M(y_i-X_i\hat\theta_i)$ where \(X_i\in\mathbb{R}^{T\times2}\) is the design matrix with first column \(\mathbf1_T\) (a \(T\)-vector of ones) and second column \(\mathbf t=(1,2,\dots,T)^\top\). Under $H_0$, the precision‐weighted pooled estimator is $\widetilde\theta_{\mathrm{WFE}}=\bigl(\sum_{i=1}^N X_i^\top M X_i/\hat\sigma_i^2\bigr)^{-1}\bigl(\sum_{i=1}^N X_i^\top M y_i/\hat\sigma_i^2\bigr)$. Each station’s weighted deviation is $\tilde d_i=(\hat\theta_i-\widetilde\theta_{\mathrm{WFE}})^\top(X_i^\top M X_i/\hat\sigma_i^2)(\hat\theta_i-\widetilde\theta_{\mathrm{WFE}})$, and the bias‐adjusted test statistic is $\widetilde D=N^{-1/2}\sum_{i=1}^N(\tilde d_i-k)/\sqrt{2k}$ with $k=2$. Under $H_0$, $\widetilde D\to N(0,1)$ and rejection at the 5\% level occurs if $\widetilde D>1.645$. For $N=1782$ stations, $\widetilde D\approx 922.6$ (p‐value~$\approx0$), rejecting the common‐trend hypothesis. Thus, there is statistically significant evidence of heterogeneous time trends in hourly demand across the 1,782 stations, so that each station’s demand evolves at its own characteristic rather than following a single, network-wide trajectory.

\begin{figure}[htbp]
  \centering
  \includegraphics[width=1\textwidth]{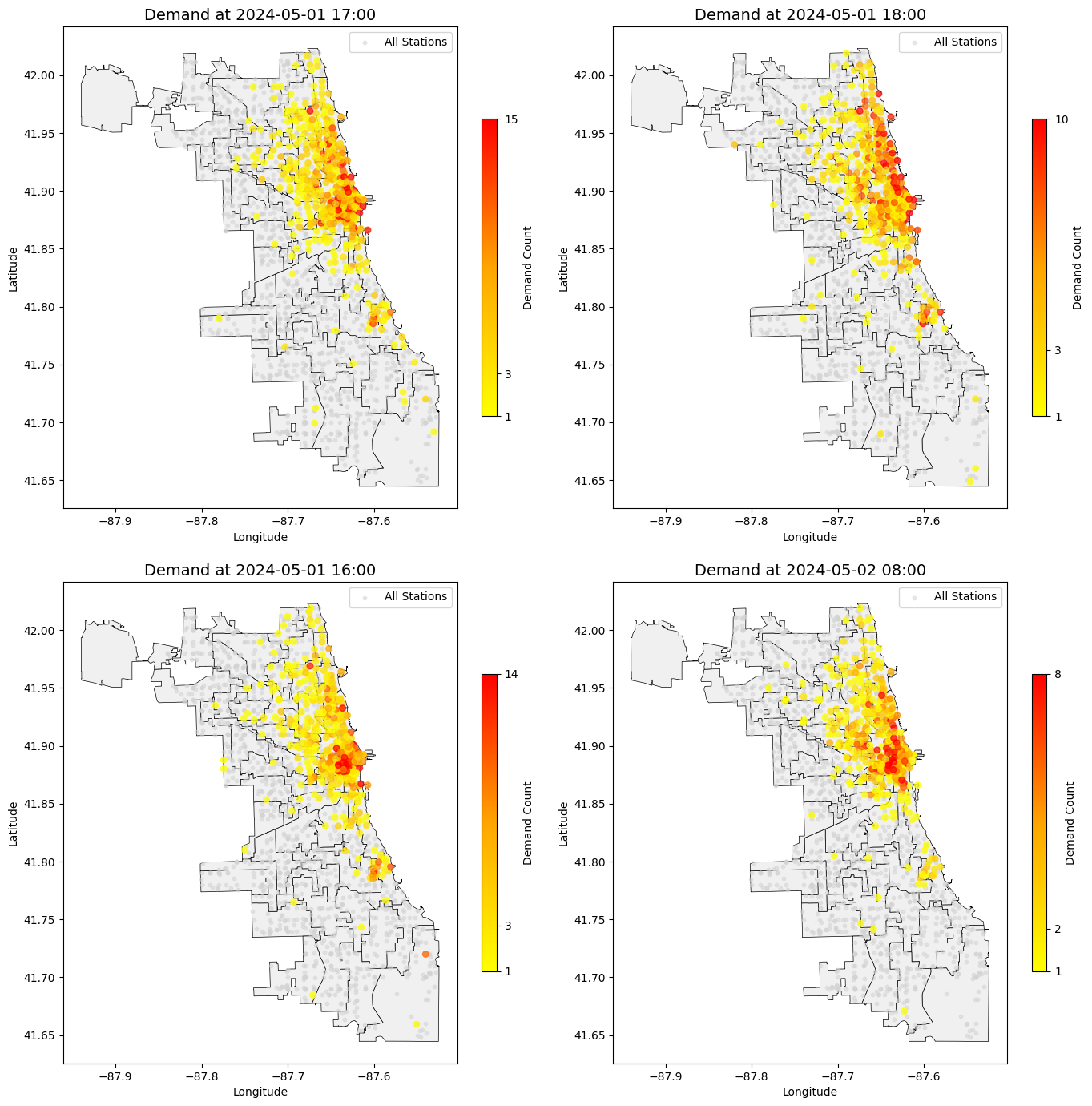}
  \caption{Spatial distribution of Divvy bike‐share demand at four peak hours}
  \label{fig:peak-hour}
\end{figure}

\autoref{fig:peak-hour} also visualizes that demand concentrates in the Loop and Near North stations during the evening rush and falls off sharply elsewhere. The maps show that demand is tightly concentrated in the downtown core at 16:00 and 17:00, but progressively disperses thereafter, becoming noticeably more diffuse by 18:00 as riders move outward. The following morning peak at 8:00 returns to a centered pattern. Overall, the spatial distribution of ridership is highly heterogeneous and evolves dynamically over time.

\subsection{Chicago American Community Survey (ACS) Data}

The ward‐level ACS data derive from the U.S. Census Bureau's American Community Survey (ACS), an ongoing household survey that produces rolling five-year estimates for social, economic, housing, and demographic characteristics.  In this application, this study uses the City of Chicago’s ACS 5-Year Data by Ward release, which reports 30 key variables of 50 wards including total population, age and sex breakdowns, race and ethnicity categories, and household income bands \citep{chicago_data_portal_acs_2025}.

Station‐hour demand records were enriched with ward‐level socioeconomic attributes via a spatial association procedure.  Each station’s coordinates were converted into map points and aligned to the Chicago 2023 ward projection.  Points were then overlaid on the official ward polygons, and the matching 30 ACS measures were transferred to each station by virtue of geographic containment.  Finally, these station‐level demographic and economic profiles were joined to the hourly demand counts using the shared station identifier, yielding a unified dataset in which every observation includes both usage intensity and its local community context. Integrating these profiles into the stations' hourly demand dataset ensures that each observation includes the full range of socioeconomic covariates necessary to evaluate the influence of local community characteristics on bike-share usage.

\autoref{tab:hourly_stacked_summary} summarizes the distributions of station‐hour demand and contextual ward‐level features across 3,553,604 observations. Demand exhibits a right‐skewed profile, with most station‐hour counts falling between one and three rides, while extreme peaks approach 141. This suggests that routine operations are centered on low to moderate volumes, punctuated by occasional surges at high-traffic locations. Demographic indicators reveal that the wards contain substantial youth cohorts (0–17 years) and young adults (18-24 years), with variability between the wards suggesting localized differences in family and student populations. Income bands cluster around middle and upper middle categories but show greater dispersion in the highest bracket, implying economic heterogeneity that could influence discretionary bike-share usage. The total population per ward remains relatively homogeneous, supporting the use of pooled models while recognizing local demand fluctuations.


\break

\subsection{Chicago Transit Authority (CTA) Data}

Infrastructure data for bus stops, rail stations, and park-and-ride facilities were obtained from the Chicago Transit Authority’s Open Data portal, using the 2022 releases for each layer \citep{chicago_transit_authority_chicago_2022}.  Divvy stations' coordinates were first deduplicated and projected to a common metric coordinate reference system.  A 100-meter-buffer was then constructed around each station to define its local catchment area.  Station-level counts of bus stops, rail stations, and park-and-ride lots were calculated by intersecting these buffers with the corresponding CTA layers, yielding three measures of multimodal accessibility for inclusion as static covariates in the clustering and forecasting analyses.

\autoref{tab:hourly_stacked_summary} shows that transit proximity measures (bus stops and rail stations within 100 meters) show minimal variance, reflecting Chicago’s uniformly dense network and indicating that micro-scale transit access alone may not differentiate demand. 

\subsection{Simulated Demand Time Series Data}

As shown in \autoref{sec:divvydata}, the slope-homogeneity test by \cite{hashem_pesaran_testing_2008} implies that there is overwhelming evidence of heterogeneous time-trends in hourly demand across the 1,782 stations. Because the real Divvy series are heterogeneous, a direct assessment of model performance under a truly homogeneous regime is not possible on the observed data alone. Thus, this study generates homogeneous synthetic datasets by using the global parameter vector $\hat\theta_{\rm global}$, fitted to the pooled average demand series, to produce 17,544 hourly trajectories for each station. This controlled setup enables a direct comparison of heterogeneous and homogeneous schemes on probabilistic forecasting of time series.

In specific, there are four candidate data‐generating processes (DGPs) to model station‐level demand (Table \ref{tab:dgps}). The first is a linear Gaussian Seasonal ARIMA of order $(p,d,q)\times(P,D,Q)_m$ \citep{box_time_2008}, whose one‐step update is given by $y_t = \mu + \sum_{i=1}^p\phi_i\,y_{t-i} + \sum_{j=1}^q\theta_j\,\varepsilon_{t-j} + \sum_{I=1}^P\Phi_I\,y_{t-Im} + \sum_{J=1}^Q\Theta_J\,\varepsilon_{t-Jm} + \varepsilon_t$, with $\varepsilon_t\sim N(0,\sigma^2)$. The second is a linear non‐Gaussian autoregression with heavy‐tailed or skewed errors \citep{aas_generalized_2006}, specified as $y_t = \mu + \sum_{\ell\in L}\phi_\ell\,y_{t-\ell} + \varepsilon_t$, $\varepsilon_t = \sigma\,z_t$, $z_t\sim D(0,1)$, where $L$ may include non‐seasonal and a 24‐hour seasonal lag and $D$ denotes a Student’s $t$ or skew-$t$ law. The third is a nonlinear Gaussian MLP‐AR($p$) \citep{zhang_forecasting_1998}: letting $x_t=[y_{t-1},\dots,y_{t-p}]^\top$, compute $h^{(1)}_t=\tanh(W^{(1)}x_t+b^{(1)})$, $h^{(2)}_t=\tanh(W^{(2)}h^{(1)}_t+b^{(2)})$, $\mu_t=W^{(3)}h^{(2)}_t+b^{(3)}$, and then $y_t=\mu_t+\varepsilon_t$, $\varepsilon_t\sim N(0,\sigma^2)$, with all weights and $\sigma^2$ estimated by minimizing mean squared error. The fourth is a linear Gaussian AR($r$) mean with a GARCH($p',q$) variance process \citep{engle_autoregressive_1982}: $y_t=\mu+\sum_{i=1}^r\phi_i\,y_{t-i}+\varepsilon_t$, $\sigma_t^2=\omega+\sum_{i=1}^{p'}\alpha_i\,\varepsilon_{t-i}^2+\sum_{j=1}^q\beta_j\,\sigma_{t-j}^2$, $\varepsilon_t\sim N(0,\sigma_t^2)$. 

\begin{table}[htp]
\centering
\scriptsize
\setlength\tabcolsep{2pt}
\renewcommand{\arraystretch}{1.1}
\caption{Data‐Generating Processes (DGPs)}
\label{tab:dgps}
\begin{tabularx}{\textwidth}{@{}p{3cm} p{5.5cm} c c c p{2.5cm}@{}}
\toprule
\textbf{Process} 
  & \textbf{One‐step update} 
  & \textbf{Linear} 
  & \textbf{Gaussian} 
  & \textbf{Variance} 
  & \textbf{Est. Param.}\\
\midrule
SARIMA $(p,d,q)(P,D,Q)_m$
  & \parbox[t]{5.5cm}{%
    $y_t=\mu+\sum_{i=1}^p\phi_i\,y_{t-i}+\sum_{j=1}^q\theta_j\,\varepsilon_{t-j}$\\
    $+\varepsilon_t+\sum_{I=1}^P\Phi_I\,y_{t-Im}+\sum_{J=1}^Q\Theta_J\,\varepsilon_{t-Jm}$
  }
  & Yes & Yes & Constant
  & \parbox[t]{3.5cm}{%
    $\{p,d,q\},\{P,D,Q\}$\\
    $\{\mu,\phi_i,\theta_j,\Phi_I,\Theta_J,\sigma^2\}$
  }\\
\addlinespace
AR$(p)$ (heavy‐tail)
  & \parbox[t]{5cm}{%
    $y_t=\mu+\sum_{\ell=1}^p\phi_\ell\,y_{t-\ell}+\varepsilon_t$\\
    $\varepsilon_t=\sigma\,z_t,\;z_t\sim D(0,1)$
  }
  & Yes & No & Constant
  & \parbox[t]{3.5cm}{%
    $p$,
    $\{\mu,\phi_\ell,\sigma,\nu,\alpha\}$
  }\\
\addlinespace
MLP‐AR$(p)$
  & \parbox[t]{5.5cm}{%
    $\displaystyle
      h_t^{(0)}=[y_{t-1},\dots,y_{t-p}]^\top,\\
      h_t^{(i)}=\tanh\!\bigl(W^{(i)}h_t^{(i-1)}+b^{(i)}\bigr)\, , i=1,2,3 \\
      y_t=h_t^{(3)}+\sigma\varepsilon_t,\;\varepsilon_t\sim N(0,1)$
  }
  & No & Yes & Constant
  & \parbox[t]{3.5cm}{%
    $p,\ \{W^{(i)},\,b^{(i)},\,\sigma\}_{i=1}^3$
  }\\
\addlinespace
AR(r) + GARCH(p,q)
  & \parbox[t]{5cm}{%
    $y_t=\mu+\sum_{i=1}^r\phi_i\,y_{t-i}+\varepsilon_t$\\
    $\sigma_t^2=\omega+\sum_{i=1}^{p}\alpha_i\,\varepsilon_{t-i}^2
                 +\sum_{j=1}^q\beta_j\,\sigma_{t-j}^2$
  }
  & Yes & Yes & Time-varying
  & \parbox[t]{3.5cm}{%
    $\{r,p,q\}$,
    $\{\mu,\phi_i,\omega,\alpha_i,\beta_j\}$
  }\\
\bottomrule
\end{tabularx}
\end{table}

Guided by the empirical Partial Autocorrelation Function (PACF), which exhibited a sharp cutoff after lag 6 and only a modest seasonal impulse at lag 24, this study restricted all AR orders to $p\le6$ and then conducted an exhaustive grid search, using AIC as the primary selection criterion and BIC as a secondary check, on the pooled Divvy average demand series. For the SARIMA$(p,d,q)\times(P,D,Q)_{24}$ class, $(p,d,q)$ was varied over $\{0,1,2,3\}\times\{0,1\}\times\{0,1,2,3\}$ and $(P,D,Q)$ over $\{0,1\}\times\{0,2\}\times\{0,1\}$. The linear non-Gaussian AR model was fitted with $p\in\{1,\dots,6\}$ under Gaussian, Student’s $t$, or skew-$t$ innovations. The MLP-AR($p$) network was evaluated across lag orders $p\in\{1,\dots,24\}$ and hidden-layer dimensions $(h_1,h_2)\in\{(10,5),(20,10),(50,25),(100,50),(200,100),(300,150)\}$. Finally, the AR($p$) plus GARCH($p',q$) class considered $p,p',q\in\{1,\dots,6\}$ with Gaussian errors. The resulting reference parameterizations identified by this search were SARIMA(3,0,3)$\times$(1,0,1)$_{24}$, AR(4) with Student’s $t$ innovations, MLP-AR(24) with hidden layers of size (100, 50), and AR(5) plus GARCH(1,6) with Gaussian errors. Under the homogeneous estimation scheme, each of the four reference models is estimated once on the pooled Divvy average series, yielding a single global coefficient vector $\hat\theta_{\rm global}$ for each template. Every station then generates its $17,544$ synthetic replicates per process by initializing at its own final observed value and iterating the common recursion and noise law under $\hat\theta_{\rm global}$. In this way, all outlets share the same parameterization, differing only in their starting point.


\begin{figure}[htbp]
  \centering
  \includegraphics[width=0.6\textwidth]{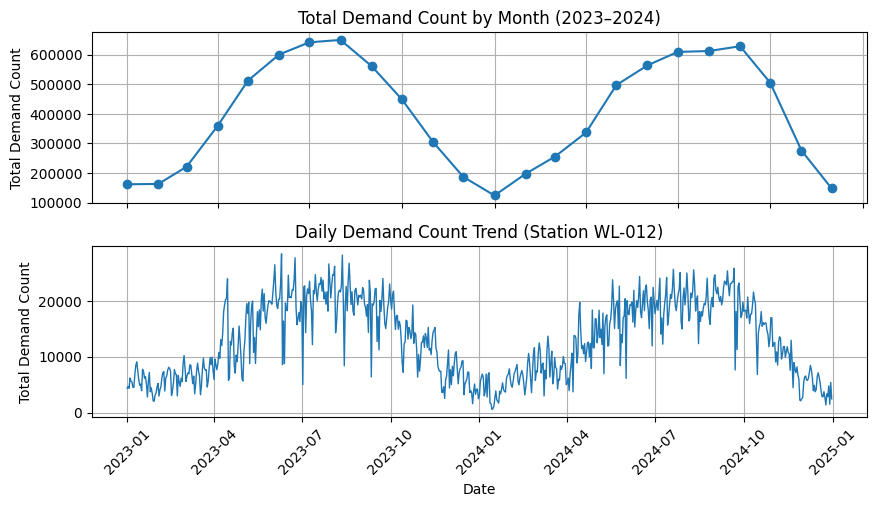}
  \caption{Daily and monthly Divvy bike-share demand time series}
  \label{fig:2023-2024}
\end{figure}

\subsection{Data Processing}

Boundary and demographic information were obtained from Chicago's ward shapefiles and the American Community Survey, and merged to create a spatially referenced layer of census attributes. CTA infrastructure data for bus stops, rail stations, and park-and-ride facilities were sourced from the 2022 Open Data releases and overlaid onto Divvy station locations to derive station-level counts of each mode within a 100-meter-buffer.  Divvy trip records for 2023 and 2024 were consolidated, cleansed, and their start times rounded to the nearest hour to compute hourly demand per station.  Finally, these hourly demand counts were combined with the static ACS and transit-access measures via spatial joins, yielding the dataset of time-indexed station's hourly demand and contextual features for clustering and forecasting.

To ensure the forecasting evaluation reflects genuine out‐of‐sample performance rather than nonstationary regime shifts, 2023 was designated as the training period and 2024 as the test period.  Aggregating demand across all stations and aligning observations by month–day–hour, the Pearson correlation between the mean hourly profiles in 2023 and 2024 was found to be $r=0.735$ ($p<0.001$), indicating a strong, statistically significant year‐over‐year reproducibility of diurnal and seasonal patterns \citep{pearson_note_1895}.

Visual inspection of the monthly total demand (\autoref{fig:2023-2024}) further confirms near‐identical peaks in mid‐summer and troughs in mid‐winter across the two years.  Moreover, no major service changes or exogenous disruptions (e.g.\ network expansions, fare adjustments, or city‐wide events) occurred between 2023 and 2024 that would undermine the comparability of the series. Taken together, these findings justify the use of the full 2023 data for model calibration and the independent 2024 data for out‐of‐sample testing.

\clearpage

\begin{table}[htbp]
  \centering
  \small
  \caption{Summary statistics of Chicago ACS by ward and hourly Divvy trips data}
  \label{tab:hourly_stacked_summary}
  \begin{subtable}[t]{\textwidth}
    \centering
    \caption{Numeric Variables}
    \begin{tabularx}{\linewidth}{l *{7}{>{\raggedleft\arraybackslash}X}}
      \toprule
      \textbf{Variable}            & \textbf{Mean} & \textbf{Std} & \textbf{Min} & \textbf{25\%} & \textbf{50\%} & \textbf{75\%} & \textbf{Max} \\
      \midrule
      Demand count                 &   2.7   &  3.1   &    1.0   &    1.0   &    2.0   &    3.0   &  141.0   \\
      Bus stops (100 m)             &   1.6   &  1.0   &    1.0   &    1.0   &    1.0   &    2.0   &    7.0   \\
      Rail stations (100 m)         &   1.0   &  0.0   &    1.0   &    1.0   &    1.0   &    1.0   &    1.0   \\
      Park-and-ride (100 m)         &   1.0   &  0.0   &    1.0   &    1.0   &    1.0   &    1.0   &    1.0   \\
      Under \$25 000               &1,141.6  &800.0   &  268.0   &  375.0   &1,019.0   &1,887.0   &3,777.0  \\
      \$25 000–\$49 999            &1,292.8  &787.8   &  278.0   &  627.0   &1,114.0   &2,082.0   &3,328.0  \\
      \$50 000–\$74 999            &1,001.0  &523.4   &  334.0   &  613.0   &  841.0   &1,468.0   &2,632.0  \\
      \$75 000–\$125 000           &1,786.2  &559.4   &  931.0   &1,490.0   &1,721.0   &1,996.0   &3,971.0  \\
      \$125 000+                   &5,882.8  &2,715.9 &  906.0   &3,281.0   &6,352.0   &7,494.0   &10,471.0 \\
      Male 0 to 17                 &4447.9   &1481.7  &2043.0    &3128.0    &4587.0    &5551.0    &8847.0 \\
      Male 18 to 24                &2657.2   &756.9   &1182.0    &2053.0    &2387.0    &3684.0    &3909.0 \\
      Male 25 to 34                &7911.8   &3271.2  &2400.0    &5469.0    &7878.0    &10147.0   &14142.0 \\
      Male 35 to 49                &6056.5   &1606.9  &3003.0    &4625.0    &5671.0    &7613.0    &8666.0 \\
      Male 50 to 64                &4206.1   &728.3   &2871.0    &3722.0    &4157.0    &4711.0    &6454.0 \\
      Male 65+                     &2647.9   &576.7   &1772.0    &2077.0    &2792.0    &3057.0    &4473.0 \\
      Female 0 to 17               &4348.2   &1434.5  &2085.0    &2934.0    &4461.0    &5518.0    &8560.0 \\
      Female 18 to 24              &3196.4   &1241.7  &1520.0    &1969.0    &2683.0    &3567.0    &5434.0 \\
      Female 25 to 34              &7893.2   &3136.0  &2498.0    &5807.0    &8078.0    &10317.0   &13644.0 \\
      Female 35 to 49              &5778.1   &1306.2  &3022.0    &4675.0    &5687.0    &6768.0    &7778.0 \\
      Female 50 to 64              &4137.7   &740.4   &2534.0    &3702.0    &4272.0    &4635.0    &6271.0 \\
      Female 65 +                  &3696.1   &1104.1  &1974.0    &2913.0    &3554.0    &4437.0    &6603.0 \\
      Total population             &56,977.3 &8,247.7 &35,489.0  &51,937.0  &55,618.0  &60,434.0  &72,572.0 \\
      White                        &30686.9  &14975.5 &449.0     &22255.0   &30916.0   &44545.0   &47407.0 \\
      African American             &13235.1  &14293.6 &701.0     &2382.0    &3848.0    &23638.0   &47497.0 \\
      Native American              &247.8    &183.7   &14.0      &101.0     &197.0     &380.0     &1322.0 \\
      Asian                        &5491.9   &4637.6  &40.0      &3435.0    &3860.0    &5356.0    &19833.0 \\
      Pacific Islander             &24.8     &28.2    &0.0       &2.0       &10.0      &41.0      &171.0 \\
      Hispanic or Latino           &8636.3   &7359.9  &633.0     &3291.0    &7082.0    &9082.0    &44965.0 \\
      \bottomrule
    \end{tabularx}
  \end{subtable} \\[1em]
  \begin{subtable}[t]{\textwidth}
    \centering
    \caption{Distribution of Divvy Trips by Weekday and Season}
    \begin{tabularx}{\linewidth}{%
        X *{2}{>{\raggedleft\arraybackslash}X}  
        @{\hspace{2cm}}                          
        X *{2}{>{\raggedleft\arraybackslash}X}  
      }
      \toprule
      \textbf{Weekday} & \textbf{Count} & \textbf{\%} 
                       & \textbf{Season} & \textbf{Count} & \textbf{\%} \\
      \midrule
      Monday      & 495,811 & 13.96 & Winter &  551,544 & 15.52 \\
      Tuesday     & 511,195 & 14.39 & Spring &  867,503 & 24.42 \\
      Wednesday   & 525,857 & 14.80 & Summer &1,151,539 & 32.43 \\
      Thursday    & 523,295 & 14.73 & Fall   &  983,018 & 27.68 \\
      Friday      & 524,989 & 14.78 &        &          &       \\
      Saturday    & 507,788 & 14.29 &        &          &       \\
      Sunday      & 464,669 & 13.08 &        &          &       \\
      \bottomrule
    \end{tabularx}
  \end{subtable}
\end{table}

\clearpage

\newpage

\section{Methodology}

\subsection{Probabilistic Forecasting Model}

LightGBM proceeds by sequentially adding trees \(f_t\) to minimize the regularized objective defined in \autoref{eq:lgbt_basic}. In \autoref{eq:lgbt_basic}, \(\ell(y_i,\hat y)\) denotes the pointwise loss for true response \(y_i\) and prediction \(\hat y\), \(\hat y_i^{(t-1)}\) is the ensemble prediction after \(t-1\) rounds, and \(\Omega(f_t)=\gamma\,T + \tfrac12\lambda\sum_{j=1}^T w_j^2\) is the complexity penalty on the new tree, where \(T\) is its number of leaves, \(w_j\) the weight of leaf \(j\), \(\gamma\) the leaf‐count penalty, and \(\lambda\) the \(L_2\) regularization parameter \citep{chen_xgboost_2016}.
\begin{equation}\label{eq:lgbt_basic}
\mathcal{L}^{(t)} = \sum_{i=1}^n \ell\bigl(y_i,\;\hat y_i^{(t-1)} + f_t(\mathbf{x}_i)\bigr) + \Omega(f_t)
\end{equation}
 
A second‐order Taylor expansion of \(\ell\) about \(\hat y_i^{(t-1)}\) yields the expanded surrogate objective in \autoref{eq:lgbt_taylor_full}, in which \(g_i = \left.\partial_{\hat y}\ell(y_i,\hat y)\right|_{\hat y=\hat y_i^{(t-1)}}\) and \(h_i = \left.\partial^2_{\hat y}\ell(y_i,\hat y)\right|_{\hat y=\hat y_i^{(t-1)}}\) are the first and second derivatives of the loss at the previous prediction. This expanded form follows by substituting the Taylor series of \(\ell\bigl(y_i,\hat y_i^{(t-1)} + f_t(\mathbf{x}_i)\bigr)\), retaining terms up to second order in \(f_t(\mathbf{x}_i)\), and including the regularization \(\Omega(f_t)\) without constant terms that do not affect optimization \citep{chen_xgboost_2016}.
\begin{equation}\label{eq:lgbt_taylor_full}
\widetilde{\mathcal{L}}^{(t)}
=
\sum_{i=1}^n \Bigl[g_i\,f_t(\mathbf{x}_i) + \tfrac12\,h_i\,f_t(\mathbf{x}_i)^2\Bigr]
+ \gamma\,T + \tfrac12\,\lambda\sum_{j=1}^T w_j^2
\end{equation}
 
LightGBM achieves superior efficiency and scalability by combining leaf‐wise tree growth, which yields deeper and more accurate splits with fewer nodes, with gradient‐based one‐side sampling, which focuses computation on high‐impact observations, and exclusive feature bundling, which compresses sparse features into compact representations \citep{ke_lightgbm_2017}. These innovations, together with histogram‐based binning, optimized memory access, and native multithreading and GPU support, enable training on large, high‐dimensional datasets orders of magnitude faster than conventional Gradient Boosted Decision Trees (GBDT) implementations without degrading accuracy.

In this study, the LightGBM framework is employed to produce probabilistic forecasts of hourly station demand by fitting quantile‐regression models at the 2.5th and 97.5th percentiles, thereby directly yielding a 95 percent prediction interval. The covariates supplied to each model consist of two lagged demand values (one‐hour and twenty‐four‐hour), calendar effects (month, day of month, and hour of day), and categorical indicators for weekday, season, and station identifier.  This combination captures both the immediate autoregressive structure of demand and its systematic temporal and station‐specific variation.

Model complexity and learning dynamics were controlled via three key hyperparameters: a learning rate of 0.05 to ensure gradual, stable convergence; 64 leaves per tree to balance model flexibility against overfitting; and 500 boosting iterations to provide sufficient ensemble size for accurate quantile estimation while controlling the computational cost.

\subsection{Clustering Approach}

In this study, K-means clustering is employed to divide $n$ stations $\{x_i\}_{i=1}^n$ into $K$ groups $C_1,\dots,C_K$ by minimizing the total within-cluster sum of squared Euclidean distances between each station’s feature vector and its assigned cluster centroid \citep{lloyd_least_1982}. The objective function for the K-means algorithm is specified in \autoref{eq:obj}.
\begin{equation}\label{eq:obj}
\min_{C_1,\dots,C_K}J
\quad\text{with}\quad
J = \sum_{k=1}^{K}\sum_{x_i \in C_k}\bigl\lVert x_i - \mu_k\bigr\rVert^2, \quad \mu_k = \frac{1}{\lvert C_k\rvert}\sum_{x_i \in C_k}x_i
\end{equation}

The clustering procedure draws on a total of 32 static covariates detailed in the Data section: ACS demographic measures and counts of nearby transportation nodes, and a set of 777 time-series features automatically extracted from each station’s hourly demand series using the EfficientFCParameters configuration of the tsfresh library, which computes a comprehensive summary statistics including autocorrelation coefficients, quantiles, entropy measures, spectral (FFT) coefficients, and other metrics \citep{christ_time_2018}.

All variables were first standardized to have mean zero and variance one to mitigate issues of high dimensionality and collinearity, and the resulting dataset was then subjected to principal component analysis with a threshold of 90 percent explained variance \citep{mackiewicz_principal_1993}. This transformation distilled approximately 809 original variables into roughly 130 orthogonal components, thereby enhancing computational tractability and improving the stability of the subsequent clustering.

Clustering was performed via the K‐Means algorithm on the PCA‐transformed data. This study evaluated candidate cluster counts $k$ in the range from 2 up to $(N_{\rm unique}-1)$, where $N_{\rm unique} \le 1782$ is the number of distinct station‐component vectors.  For each $k$, sum of squares (WSS) and the mean silhouette score are computed.  The optimal $k$ was identified by applying a knee‐point detection algorithm (KneeLocator) to the WSS curve and cross‐checked via silhouette values \citep{satopaa_finding_2011}.

\subsection{Probabilistic Forecast Metrics}

The prediction interval coverage probability (PICP) measures the empirical frequency with which observed values fall within the nominal \(95\%\) prediction intervals.  It is defined in \autoref{eq:picp}, where \(\hat y_t^{\mathrm{lower}}\) and \(\hat y_t^{\mathrm{upper}}\) denote the 2.5th and 97.5th percentile forecasts, respectively.  PICP directly evaluates interval calibration and was chosen to verify that the nominal coverage level is met in practice.
\begin{equation}\label{eq:picp}
\mathrm{PICP}
=
\frac{1}{N}\sum_{t=1}^N
\mathbf{1}\bigl(y_t \in [\,\hat y_t^{\mathrm{lower}},\,\hat y_t^{\mathrm{upper}}]\bigr).
\end{equation}

The prediction interval normalized average width (PINAW) quantifies the average sharpness of the \(95\%\) intervals, normalized by the observed range, and is defined in \autoref{eq:pinaw}.  By penalizing excessively wide intervals, PINAW complements PICP and promotes parsimonious uncertainty quantification.
\begin{equation}\label{eq:pinaw}
\mathrm{PINAW}
=
\frac{1}{N}\sum_{t=1}^N
\frac{\hat y_t^{\mathrm{upper}} - \hat y_t^{\mathrm{lower}}}
{\max_{1\le u\le N}y_u - \min_{1\le u\le N}y_u}.
\end{equation}

The median mean squared error (MSE\(_\mathrm{median}\)) in \autoref{eq:mse_med} evaluates the accuracy of the 50th‐percentile (median) forecast.  This metric was selected to assess point‐forecast performance in a manner directly comparable to conventional mean‐squared‐error benchmarks.
\begin{equation}\label{eq:mse_med}
\mathrm{MSE}_{\mathrm{median}}
=
\frac{1}{N}\sum_{t=1}^N
\bigl(y_t - \hat y_t^{\mathrm{median}}\bigr)^2.
\end{equation}

Forecast performance was assessed using PICP to evaluate interval calibration and PINAW to measure interval sharpness, alongside median MSE for point‐forecast accuracy. These metrics provide a comprehensive evaluation of probabilistic forecasting models.

\begin{figure}[htbp]
  \centering
  \includegraphics[width=1\textwidth]{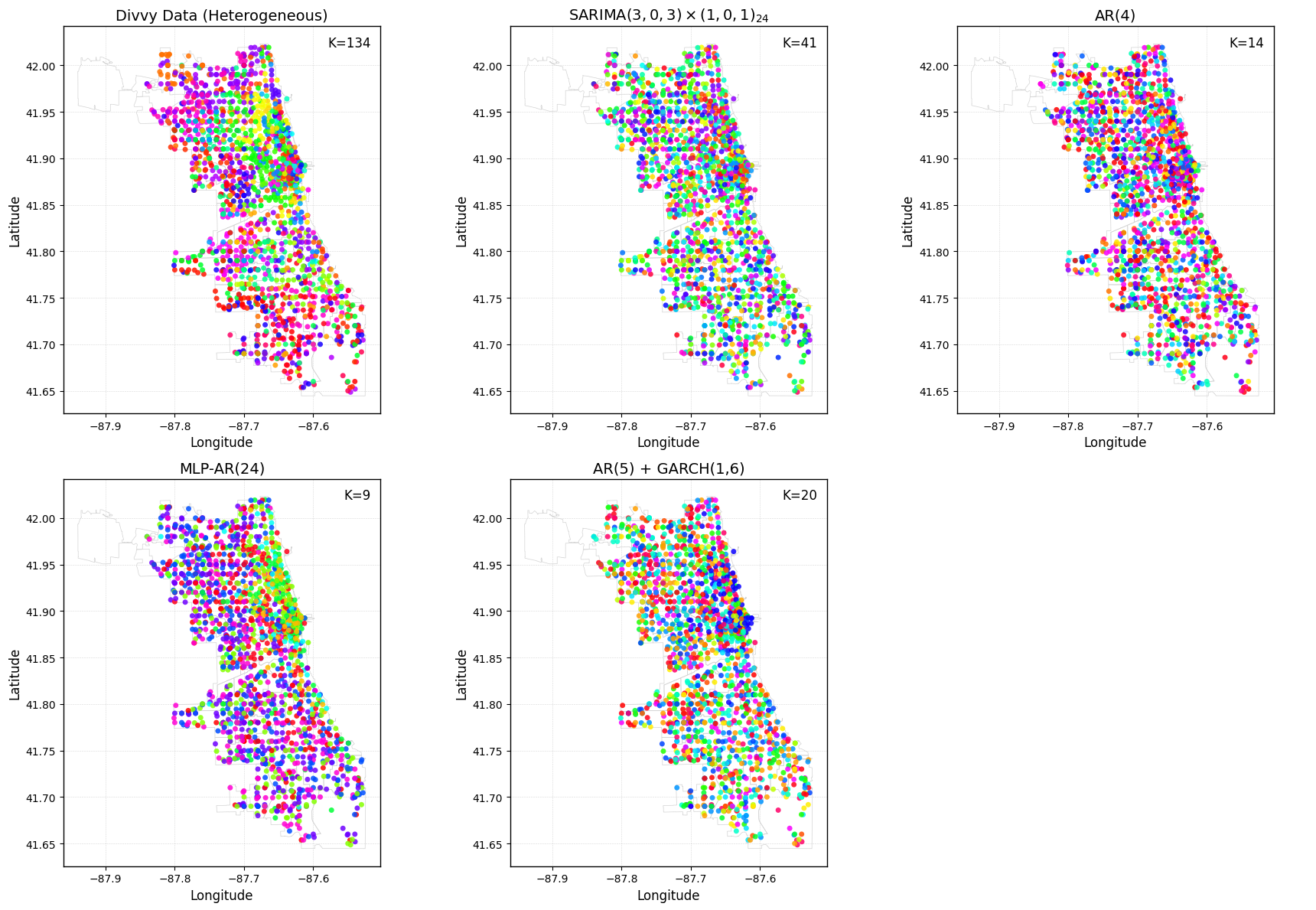}
  \caption{Spatial clustering comparison of Divvy bike demand time series across DGPs}
  \label{fig:clusters}
\end{figure}

\newpage

\section{Results and Discussion}

\subsection{Station's Demand Time Series Clustering}

This study clustered 1,782 stations under 5 data generating processes (DGPs) and found the following grouping structures in ascending order of cluster count. The MLP-AR model with 24 lags and hidden layers of size 100 and 50 yielded 9 clusters averaging 198 stations each. The AR of order 4 with Student’s t innovations produced 14 clusters with a mean size of 127. The AR(5) plus GARCH(1,6) simulation formed 20 clusters averaging 89 stations. The SARIMA(3,0,3)$\times$(1,0,1) at lag 24 simulation returned 41 clusters with an average of 43 stations. Finally, the real Divvy data split into 134 clusters of mean size 13 stations.

Starting with the smallest cluster count, the MLP-AR simulation generated remarkably uniform series behavior under these settings, so only a few very large groups were needed to capture the shared dynamics. Introducing Student’s t innovations in the AR(4) model added heavy tail variation, breaking the data into more clusters but still yielding large groupings. Adding conditional heteroskedasticity in the AR(5) plus GARCH(1,6) process increased temporal complexity further, creating even more distinct regimes and thus more clusters of moderate size. The SARIMA model’s combination of seasonal and nonseasonal components produced diverse periodic patterns, resulting in a higher number of smaller clusters. In contrast, the real-world data contain myriad unmodeled influences and bespoke station behaviors, which fragment the series into many small clusters. Together, these results illustrate how increasing the richness of temporal dependence in simulated data gradually fragments station groupings from a few large clusters to many small ones, with real data exhibiting the greatest heterogeneity.

\autoref{fig:clusters} displays the spatial clustering of Divvy stations based on read Divvy Bike Data and data simulated under SARIMA(3,0,3)$\times$(1,0,1)$_{24}$, AR(4) with Student’s $t$ innovations, MLP‐AR(24) with hidden layers of size (100, 50), and AR(5) plus GARCH(1,6), where each colored point represents a station’s assignment to one of the $K$ clusters (value of $K$ annotated in each panel), thus demonstrating how the choice of simulation model affects the spatial coherence and resolution of the resulting cluster solutions.

\subsection{Global, Cluster and Local Probabilistic Forecasting Comparison}

In the evaluation across five distinct data generating processes the global LightGBM model exhibited the most conservative calibration. It achieved prediction interval coverage probabilities of 0.9885 for the Divvy bike share data, 0.9469 under the SARIMA process, 0.9483 for the autoregressive heavy tail series, 0.9500 for the multilayer perceptron autoregressive simulation and 0.9486 for the autoregressive plus conditional heteroskedasticity scenario. By comparison the cluster-level model yielded coverage probabilities of 0.9566, 0.9338, 0.9432, 0.9472 and 0.9286 respectively, indicating slight undercoverage in the more volatile settings. The station-level model exhibited the greatest deviation from the nominal 95\% coverage. In the Divvy application it over-covered with a PICP of 0.9814, exceeding the cluster-level result but remaining below the global model, while in the four synthetic scenarios it under-covered severely, with PICP values between 0.8338 and 0.8538, falling well below both global and cluster-level performance. These findings can be explained by the inclusion of the station identifier as a categorical covariate in both the global and cluster-level LightGBM frameworks. The inclusion of the station identifier as a categorical covariate in both the global and cluster-level LightGBM frameworks facilitates station-specific adjustments within each pooled model, which enables the global and cluster-level approaches to maintain superior empirical coverage relative to the station-level model.

In the assessment of normalized interval widths the global LightGBM model consistently produced the narrowest prediction intervals across all data generating processes. This outcome aligns with prior work showing that complete pooling minimizes estimation variance and yields sharper probabilistic forecasts in high dimensional settings \citep{johnson_bayes_2022}. In the Divvy application the global PINAW was approximately 0.0061 compared to 0.0364 for the station-level model and 0.1006 for the cluster-level model. In the SARIMA and MLP-AR processes the PINAW ordering followed global less than local less than cluster, whereas in the autoregressive heavy tail and autoregressive plus conditional heteroskedasticity scenarios the cluster-level model occupied an intermediate position, yielding the ordering global less than cluster less than local. By pooling data from all stations, the global model leverages a large effective sample size to stabilize variance estimates, resulting in relatively narrow absolute prediction intervals. This reversal in more heterogeneous contexts arises because cluster-level pooling only reduces estimation variance when series within each group share truly similar dynamics. If clusters retain substantial heterogeneity the quantile regression model must accommodate divergent patterns across series, inflating residual variance and widening prediction intervals beyond those from station-level fits. Prior research on structural ensemble regression has demonstrated that the number and quality of clusters critically affects forecasting performance since suboptimal partitions can create unbalanced subsets that degrade accuracy \citep{kontogiannis_structural_2022}. In this study, the K-means clustering applied to the Divvy data and to the SARIMA and MLP-AR processes appears to have formed groups with sufficient residual within cluster variation that the benefits of partial pooling were outweighed by increased dispersion within those clusters.

\begin{table}[ht]
\centering
\caption{Prediction‐Interval performance of a global LightGBM across DGPs}
\label{tab:global_lightgbm_performance}
\begin{tabular}{lrrr}
\toprule
Data Generating Process & PICP & PINAW & MSE$_\mathrm{median}$ \\
\midrule
Divvy Bike Data & \textbf{0.98847} & 0.00608 & 0.82696 \\
SARIMA(3,0,3)\,$\times$\,(1,0,1)$_{24}$ & 0.94689 & 0.15388 & 0.00910 \\
AR(4) (heavy-tail)  & 0.94831 & 0.02609 & 0.02657 \\
MLP-AR(24)  & 0.94998 & 0.04593 & \textbf{0.00482} \\
AR(5) plus GARCH(1,6) & 0.94859 & \textbf{0.01158} & 1.78503 \\
\bottomrule
\end{tabular}
\end{table}

Across all synthetic data‐generating processes the fully pooled global LightGBM model attains the lowest median‐forecast mean squared error, evidencing its ability to learn common temporal structures from the complete cross‐sectional sample. In the Divvy bike‐share application the station‐level model marginally outperforms the global estimator, reflecting the dominance of idiosyncratic demand profiles when sufficient local data are available. Cluster‐level pooling, by contrast, yields the highest median‐forecast errors in every scenario. This ordering, where global and local are roughly equivalent and both lie below cluster, underscores that partial pooling can sacrifice local adaptivity without fully capturing global structure. In effect, clustering reduces variance compared with purely local fits but introduces bias when groups remain heterogeneous, resulting in median forecasts that are worse than those from either a fully pooled model or purely local models.

\subsection{Prediction Interval Performance on Homogeneous and Heterogeneous Data}

Under the homogeneous simulations, all series generated by the same underlying process, the global LightGBM model naturally leverages the full cross‐sectional sample to estimate quantiles, yielding PICP tightly clustered around the nominal 95\% level.  By contrast, the cluster-level and station-level models substantially under-cover in these homogeneous cases, with mean PICPs falling into the low-90 percent and mid-80 percent ranges respectively. This pattern reflects the global model’s ability to leverage identical generating processes across all series, whereas clustering introduces residual heterogeneity and purely local fits suffer from extreme estimation variance on small samples. In comparison, in the Divvy data, the global model attains the highest empirical coverage (PICP = 0.9885), followed by the station-level model (PICP = 0.9814) and then the cluster-level model (PICP = 0.9566). The global estimator achieves superior coverage by pooling all observations and including the station identifier as a categorical covariate, which aligns its predictive quantiles with each station’s baseline while drawing on the full cross-sectional sample to encompass almost every outcome. In contrast, the cluster-level model groups stations into subpopulations that still exhibit residual heterogeneity. This misalignment between pooled quantile estimates and individual series distributions leads to the lowest coverage among the three schemes. Consequently, partial pooling via clustering can underperform both full pooling with station adjustments and fully local estimation in the presence of pronounced demand heterogeneity.

\begin{table}[ht]
  \centering
  \small
  \renewcommand{\arraystretch}{1.2}
  \caption{Prediction‐Interval performance of cluster‐level LightGBM across DGPs}
  \label{tab:cluster-model}
  \resizebox{\linewidth}{!}{%
    \begin{tabular}{lccccc}
      \toprule
      \multirow{2}{*}{\bfseries Statistic} 
        & \multicolumn{5}{c}{\bfseries Data‐Generating Process (DGPs)} \\
      \cmidrule(l){2-6}
        & {\footnotesize Divvy Data} 
        & {\footnotesize SARIMA(3,0,3)$\times$(1,0,1)$_{24}$} 
        & {\footnotesize AR(4) (heavy-tail)} 
        & {\footnotesize MLP-AR(24)} 
        & {\footnotesize AR(5) + GARCH(1,6)} \\
      \midrule
      \textbf{Num of Clusters} & 134 & 41 & 14 & 9 & 20 \\
      \multicolumn{6}{l}{\textbf{PICP}} \\
      Mean   & \textbf{0.95661} & 0.93380 & 0.94316 & 0.94724 & 0.92860 \\
      Min    & 0.86262 & 0.85280 & 0.91702 & \textbf{0.94543} & 0.84085 \\
      Med    & \textbf{0.95162} & 0.93860 & 0.94454 & 0.94747 & 0.94456 \\
      Max    & \textbf{0.99968} & 0.94528 & 0.94864 & 0.94964 & 0.94797 \\
      Std    & 0.02916 & 0.01645 & 0.00777 & \textbf{0.00118} & 0.03342 \\
      \addlinespace

      \multicolumn{6}{l}{\textbf{PINAW}} \\
      Mean   & 0.10055 & 0.17292 & 0.04599 & 0.05816 & \textbf{0.03915} \\
      Min    & \textbf{0.00000} & 0.15740 & 0.02953 & 0.05154 & 0.01188 \\
      Med    & 0.10858 & 0.17198 & 0.04430 & 0.05270 & \textbf{0.02771} \\
      Max    & 0.24866 & 0.18955 & \textbf{0.08269} & 0.09181 & 0.10250 \\
      Std    & 0.06961 & \textbf{0.00747} & 0.01475 & 0.01317 & 0.02480 \\
      \addlinespace

      \multicolumn{6}{l}{\textbf{MSE$_\mathrm{median}$}} \\
      Mean   & 3.61862 & 0.00929 & 0.02662 & \textbf{0.00521} & 1.80727 \\
      Min    & \textbf{0.00050} & 0.00914 & 0.02602 & 0.00390 & 1.61401 \\
      Med    & 2.80769 & 0.00925 & 0.02657 & \textbf{0.00426} & 1.77739 \\
      Max    & 32.45683 & \textbf{0.00991} & 0.02748 & 0.01163 & 2.40448 \\
      Std    & 4.51740 & \textbf{0.00016} & 0.00043 & 0.00246 & 0.17149 \\
      \bottomrule
    \end{tabular}%
  }
\end{table}

In comparing PINAW between homogeneous and heterogeneous regimes, clear distinctions emerge. Because PINAW divides by the overall data range, pooling heterogeneous series increases the denominator and lowers PINAW, while using a single series with its narrower range raises PINAW for the same absolute interval widths. Under the homogeneous simulations all series follow identical data generating processes, and the global model’s PINAW ranges modestly with process complexity, typically exceeding narrowness by only a few hundredths when moving to cluster or station models. In these settings clustering and local fitting each incur only a small increase in interval width because no true heterogeneity must be reconciled. By contrast, the heterogeneous Divvy data exhibit pronounced station‐to‐station variation, and complete pooling compresses intervals to a PINAW of roughly 0.006 by averaging over all series. The station model, which treats each series independently, yields a moderate PINAW near 0.036, reflecting high local variance, while the cluster model must negotiate residual variability within imperfectly formed groups and therefore produces the widest intervals at about 0.101. Thus, when series share a common structure pooling advantages diminish, but in the presence of strong heterogeneity full pooling dramatically sharpens intervals at the expense of bias, whereas partial and local approaches progressively sacrifice sharpness to respect individual series dynamics.

\begin{table}[ht]
  \centering
  \small
  \renewcommand{\arraystretch}{1.2}
  \caption{Prediction‐Interval performance of station‐level LightGBM across DGPs}
  \label{tab:local-model}
  \resizebox{\linewidth}{!}{%
    \begin{tabular}{lccccc}
      \toprule
      \multirow{2}{*}{\bfseries Statistic} 
        & \multicolumn{5}{c}{\bfseries Data‐Generating Process (DGPs)} \\
      \cmidrule(l){2-6}
        & {\footnotesize Divvy Data} 
        & {\footnotesize SARIMA(3,0,3)$\times$(1,0,1)$_{24}$} 
        & {\footnotesize AR(4) (heavy-tail)} 
        & {\footnotesize MLP-AR(24)} 
        & {\footnotesize AR(5) + GARCH(1,6)} \\
      \midrule

      \multicolumn{6}{l}{\textbf{PICP}} \\
      Mean   & \textbf{0.98141} & 0.83379 & 0.84967 & 0.84972 & 0.85377 \\
      Min    & 0.81956 & 0.79713 & 0.82525 & 0.81478 & \textbf{0.82855} \\
      Med    & \textbf{0.99636} & 0.83407 & 0.84990 & 0.84990 & 0.85383 \\
      Max    & \textbf{1.00000} & 0.86532 & 0.87523 & 0.88035 & 0.88058 \\
      Std    & 0.02506 & 0.01005 & 0.00782 & 0.00903 & \textbf{0.00752} \\
      \addlinespace

      \multicolumn{6}{l}{\textbf{PINAW}} \\
      Mean   & \textbf{0.03638} & 0.16951 & 0.09336 & 0.04759 & 0.07526 \\
      Min    & \textbf{0.00000} & 0.12569 & 0.02781 & 0.03853 & 0.01156 \\
      Med    & \textbf{0.01048} & 0.16968 & 0.09483 & 0.04137 & 0.07559 \\
      Max    & 0.25275 & 0.21279 & 0.13373 & \textbf{0.09341} & 0.13429 \\
      Std    & 0.05248 & \textbf{0.01185} & 0.01654 & 0.01281 & 0.01897 \\
      \addlinespace

      \multicolumn{6}{l}{\textbf{MSE$_\mathrm{med}$}} \\
      Mean   & 0.51728 & 0.01029 & 0.02857 & \textbf{0.00514} & 1.89352 \\
      Min    & \textbf{0.00000} & 0.00928 & 0.02371 & 0.00380 & 1.34038 \\
      Med    & 0.00546 & 0.01027 & 0.02767 & \textbf{0.00423} & 1.74077 \\
      Max    & 32.45683 & \textbf{0.01166} & 0.10016 & 0.01925 & 43.09306 \\
      Std    & 1.68537 & \textbf{0.00038} & 0.00441 & 0.00226 & 1.11550 \\
      \bottomrule
    \end{tabular}%
  }
\end{table}

For homogeneous processes the global estimator consistently attains the lowest median-forecast mean squared error, reflecting its efficient use of common structural information. Cluster-level and station-level fits yield modestly higher errors, as clustering misassignments introduce bias and local sample limitations inflate variance. In contrast, the heterogeneous Divvy scenario the station-level model delivers the most accurate point forecasts, achieving a median mean squared error of approximately 0.005 by tailoring its estimates to each station’s unique demand pattern. The global model, which pools all stations while adjusting for station identity, incurs a substantially higher error of roughly 0.827. The cluster-level model performs worst, with an error near 3.619, because grouping stations into imperfect clusters mixes dissimilar demand profiles and introduces bias that outweighs any variance reduction. Collectively, these findings demonstrate that the relative performance of pooling strategies critically depends on the underlying degree of cross-series homogeneity.

\subsection{Effects of Distributional Properties}

LightGBM’s built-in quantile-regression objective imposes no distributional or functional form on the data-generating process \citep{ke_lightgbm_2017}. As a tree-based ensemble it nonparametrically learns conditional quantiles from the features and target without assuming Gaussianity, linearity, or any specific tail behavior. Consequently it can accommodate normal, skewed, heavy-tailed or even multimodal distributions and capture both simple and complex nonlinear dependencies. However, in the presence of conditional heteroskedasticity the absence of dedicated variance-related features prevents the model from adjusting its interval estimates to local volatility regimes, leading to systematically biased quantile forecasts and degraded calibration. The results indicate that linearity, Gaussian innovations, and constant variance exert no systematic impact on the comparative efficacy of global, cluster‐level, and station‐level quantile‐regression models.

\newpage

\section{Conclusions}

This study evaluated three LightGBM probabilistic forecasting strategies, including global pooling, cluster-level pooling, and station-level modeling, across five data generating scenarios ranging from fully homogeneous simulations to the highly heterogeneous real Divvy bike-share demand data for 2023 through 2024. Clustering was performed using the K-means algorithm on principal component analysis (PCA) transformed covariates, which comprised extracted time series features, counts of nearby transportation nodes, and local demographic measures. Forecast performance was assessed by prediction interval coverage probability (PICP), normalized interval width (PINAW) and median mean squared error (MSE) of median forecasts.

This study illustrates the effectiveness of probabilistic forecasting using global LightGBM models that incorporate station identifiers, effectively combining broad data pooling with local customization. Empirical results across five diverse scenarios confirm that the global model consistently achieves the highest prediction interval coverage probabilities (PICP of 98.85\% for Divvy), narrowest intervals (PINAW of 0.006 for Divvy), and lowest median forecast errors in most cases. Conversely, clustering-based methods frequently underperform, with improper cluster formation significantly diminishing accuracy (PICP of 95.66\% for Divvy), evidenced by substantially wider prediction intervals (PINAW of 0.101 for Divvy) and higher median forecast errors, while also incurring higher computational costs due to repeated model initialization and inefficient use of computational resources. Station-level modeling, despite substantial under-coverage (PICP around 85\%) in simulated scenarios, demonstrated improved accuracy on heterogeneous real-world data (PICP of 98.14\% for Divvy), capturing local station-specific demand patterns effectively. Overall, explicit modeling of station identity within a globally pooled framework consistently delivered superior forecasting performance, emphasizing the importance of aligning modeling strategies closely with underlying data characteristics and structural assumptions.

This research demonstrates that integrating station identifiers into a unified LightGBM framework allows the model to benefit from extensive pooled data while still adapting to individual station behaviors. In contrast, approaches based on unsupervised clustering often struggle since misgrouped stations dilute important local variations, leading to poorer coverage and bloated interval widths. Purely station-specific models can excel when demand patterns vary greatly across locations, but they suffer from unreliable intervals and higher errors when data are scarce or too uniform. By contrast, the globally trained model with explicit station adjustments strikes the optimal balance, delivering consistently tight, reliable forecasts without the pitfalls of either extreme.

While the findings from this study highlight the advantages of global LightGBM models with station identifiers, several important limitations must be acknowledged. First, using simulated data generating processes for controlled comparison may not capture the full complexity of real-world demand drivers such as weather, special events, and service disruptions, and the fixed hyperparameter settings for LightGBM including learning rate, lag structure, and cluster count selection may not generalize beyond the Divvy context. Second, the clustering relies solely on static station features and principal component analysis reduced representations and does not incorporate dynamic external variables such as weather or transit delays, potentially overlooking temporal shifts in station behavior. Third, K-means clustering assumes spherical groups and equal variance, which may fail to reflect complex spatial temporal dependencies or non Euclidean similarity among stations. Finally, all of the data generating processes assume stationarity during the training and test periods, even though real bike-share systems often exhibit evolving usage patterns over seasons or years, indicating a need for ongoing model retraining and concept drift management.

Building on these limitations, several directions for future work emerge. One avenue is to expand the set of forecasting models under comparison. Statistical approaches such as Exponential Smoothing models like ETS may perform differently from machine learning methods when station level indicators are not available, while models like Autoregressive Integrated Moving Average with exogenous inputs (ARIMAX) that can include station identifiers would provide a classical benchmark against which to assess the benefits observed in LightGBM and LSTM. Another opportunity lies in exploring alternative clustering techniques. In particular, spatially constrained hierarchical methods such as SKATER could offer practical advantages in applications where geographic continuity is critical, for example in crowdsourced delivery or area zoning. Further research should also investigate how to evaluate and optimize cluster quality so that grouping aligns more closely with latent demand structures. Finally, future simulation studies could impose integer constraints on synthetic demand values to more closely approximate real-world settings, where demand must be represented in whole units.

\newpage





\newpage


\bibliographystyle{plain}   
\bibliography{references}   


\end{document}